\documentclass[default]{sn-jnl}
\jyear{2022}%

\usepackage[utf8]{inputenc}

\usepackage{graphicx}
\usepackage{listings}
\usepackage{xcolor}
\usepackage{ifthen}
\usepackage{xspace}
\usepackage{amssymb}
\usepackage{hyperref}
\usepackage{float}
\usepackage{subcaption}
\usepackage{enumitem}

\definecolor{verylightgray}{rgb}{0.95, 0.95, 0.95}

\usepackage{xcolor}
\newboolean{showcomments}
\setboolean{showcomments}{true}
\ifthenelse{\boolean{showcomments}}
  {\newcommand{\bnote}[2]{
	\fbox{\bfseries\sffamily\scriptsize#1}
    {\sf\small$\blacktriangleright$\textit{#2}$\blacktriangleleft$}
   }
   
  }
  {\newcommand{\bnote}[2]{}
   
  }

\graphicspath{{figures/}}

\newcommand{\commented}[1]{}

\newcommand{\eg}{\emph{e.g.,}\xspace}
\newcommand{\ie}{\emph{i.e.,}\xspace}

\usepackage{url}            
\makeatletter
\def\url@leostyle{%
  \@ifundefined{selectfont}{\def\UrlFont{\sf}}{\def\UrlFont{\small\sffamily}}}
\makeatother
\begin{document}

\newcommand{\ascode}[1]{\textsf{#1}}
\newcommand{\rulename}[1]{\textsf{#1}}
\newcommand{\arr}{=$>$~}
\newcommand{\etAl}[2]{\emph{#1} et al~\cite{#2}}

\title[Interactive, Iterative, Tooled, Rule-Based Migration]{Interactive, Iterative, Tooled, Rule-Based Migration of Microsoft Access to Web Technologies}

\author*[1,2]{\fnm{Santiago} \sur{Bragagnolo}}\email{santiago.bragagnolo@berger-levrault.com}

\author[1]{\fnm{Nicolas} \sur{Anquetil}}\email{nicolas.anquetil@univ-lille.fr}

\author[1]{\fnm{Stephane} \sur{Ducasse}}\email{stephane.ducasse@inria.fr}

\author[2]{\fnm{Abderrahmane} \sur{Seriai}}\email{abderrahmane.seriai@berger-levrault.com}

\author[2]{\fnm{Mustapha} \sur{Derras}}\email{mustapha.derras@berger-levrault.com}

\affil*[1]{ \orgname{Université de Lille, CNRS, Inria, Centrale Lille, UMR 9189 – CRIStAL}, \orgaddress{\country{France}}}
\affil*[2]{ \orgname{Berger-Levrault}, \orgaddress{\country{France}}}


\abstract{
	In the context of a collaboration with Berger-Levrault, an IT company producing information systems, we are working on migrating Microsoft Access monolithic applications to the web front-end and microservices back-end.
	Like in most software migrations, developers must learn the target technology, and they will be in charge of the evolution of the migrated system in the future.
	To respond to this problem, we propose the developers take over the migration project. 
	To enable the developers to drive the migration to the target systems, we propose an Interactive, Iterative, Tooled, Rule-Based Migration approach.

	The contributions of this article are 
	(i) an iterative, interactive process to language, library, GUI and architectural migration;
	(ii) proposal of a set of artefacts required to support such an approach; 
	(iii) three different validations of the approach: 
		(a) library and paradigm use migration to Java and Pharo, 
		(b) tables and queries migration to Java and Typescript, 
		(c)  form migration to Java Springboot and Typescript Angular. 

}

\maketitle

\section{Introduction}
\label{sec:introduction}

In the context of a collaboration with Berger-Levrault, an IT company, we are migrating Microsoft Access monolithic applications to  Typescript and Angular front end; Java and SpringBoot back end.

The migration constraints are those often encountered in the industry: The software quality is uncertain, development must continue during the migration, resources are limited, the desired target architecture is sketchy, developers are yet to master the target technology fully, and they will be in charge of the evolution of the migrated system in the future.

The last point is considered vital.
Despite the changes in architecture and technology, the development team must be able to retain or migrate its knowledge of the application to continue evolving it in the future.
To ensure this, we propose that the development team be the driving force behind the migration, choosing what to migrate, where and how to migrate.
Yet, as already stated, they are not specialists in the target technologies (Typescript and Angular) and software migration.
We propose an interactive, iterative/incremental and tooled migration process to allow them to achieve this migration.

\begin{description}
	\item[Abstraction.] A tooled approach allows software engineers to concentrate on what they want to achieve without considering how to achieve it. For example, they can transform a function into a method without worrying about transformation rules or the syntax of methods in the target language.

	\item[Low barrier.] An interactive approach enables developers to point to a software artefact, \eg a function, and decide where to migrate it, \eg a class, without worrying \emph{how to} do it. 
	The fact that the approach is based on rules reduces the need for migration experts, as experts have to put their knowledge into reusable fine-tuned rules. 

	\item[Gradual learning.] An iterative and incremental approach allows developers to migrate screens or functionalities one by one. This, plus the feedback provided by a tooled approach, allows one to gradually get acquainted with the migrated solution, its new architecture, and its new technology. 

\end{description}

In this paper, we present an approach and a tool to realise this interactive and iterative migration.
The paper is organised as follows:
\begin{itemize}
	\item In \autoref{sec:overview}, we give a first high-level view of the approach and its main components;
	
	\item In \autoref{sec:directive-rules}, we present the different elements of our migration infrastructure, which make our approach possible: models, directives, mappings and rules. 

	\item In \autoref{sec:approach-in-action}, we present the engine which puts all the artefacts together and enables the application of directives.
	
	\item In \autoref{sec:valid}, we validate the approach by presenting results on three different kinds of migration.
	\item In \autoref{sec:discussion}, we discuss different aspects and decisions of our approach.
	\item In \autoref{sec:future-conclusion}, we propose future work and conclude.
\end{itemize}

\section{Overview}
\label{sec:overview}

We  introduce the challenges and the expectations of our migration approach in \autoref{sec:challenges}, 
and we sketch the proposed process and its different parts in \autoref{sec:overview-process}.

\subsection{Challenges \& Requirements}
\label{sec:challenges}

The migration project from a monolithic Microsoft Access application to a front-end (Typescript + Angular) and back-end (Java + SpringBoot) application entails the following challenges:
\begin{description}
	\item[Programming language migration:] From Visual Basic for Applications (VBA) to multiple targets: HTML, TypeScript, and Java;

	\item[Paradigm migration:] From procedural paradigm (VBA) to object-oriented paradigm (Java and Typescript).
		Note that we do not pay close attention to this challenge in the paper.

	\item[Library migration:] The migration includes migrating libraries, 
		\eg The ``long'' type in Microsoft Access must be migrated to ``BigInteger'' in Java or ``bigint'' in Typescript.
		``MsgBox'' function usages are transformed to ``alert'' in Typescript or a Log4j ``logger'' in Java;

	\item[Architectural Migrations:] 
		The migration includes the splitting of a single stand-alone application into a front-end and a back-end application;
		it also includes splitting both front-end and back-end concerns into multiple micro-service applications.
		
	\item[GUI migration.]  From desktop GUI to web GUI;

	\item[Migrating the expertise:] As much as possible (considering the architectural changes), artefacts (classes, functions, forms) should be migrated so that developers recognise them afterwards and know where to find them;

	\item[Integrated in the typical lifecycle:]  The source and target projects are expected to have their lifecycles so that new functionalities and bug fixes can be processed with the slightest disturbances and shipped and deployed independently.

\end{description}

\subsection{An Iterative \& Interactive Process for software migration}
\label{sec:overview-process}

\begin{figure}[htbp]
	\centering
	\includegraphics[scale=0.30]{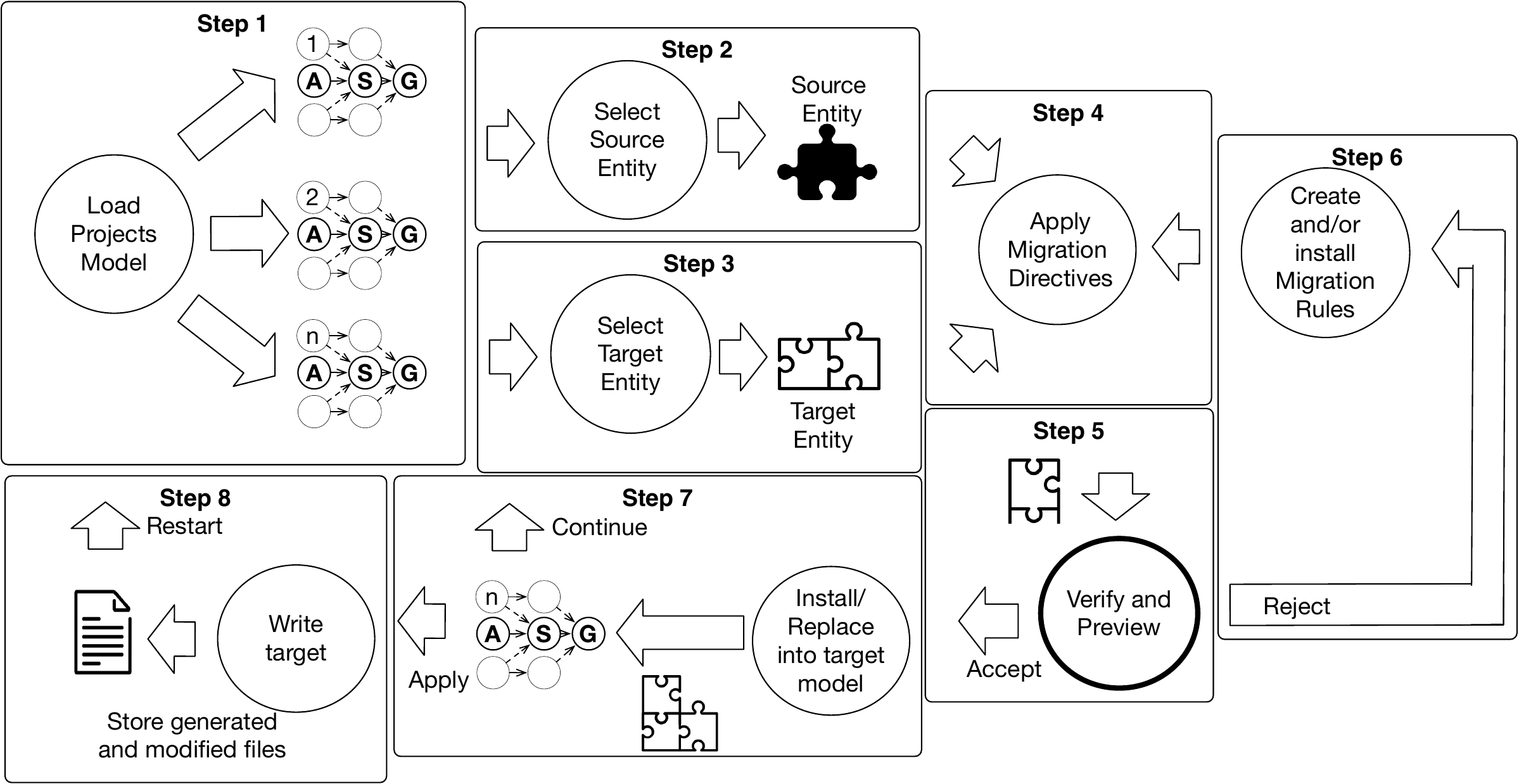}
	\caption{\label{fig:migration-process} Iterative Interactive Migration Process Steps.}
\end{figure}

The process that we propose is depicted in \autoref{fig:migration-process}; it has the following steps:
\begin{enumerate}[label=\textbf{Step \arabic*.},align=left]
	\item The migration tool loads the source and target applications and creates a model for each of them.
	      These models are presented to the software developers in the tool as a tree of software artefacts and matching source code (see, for example, Figure~\ref{fig:srcTargetModels}).

	      \begin{figure}[htbp]
		      \centering
		      \includegraphics[scale=0.4]{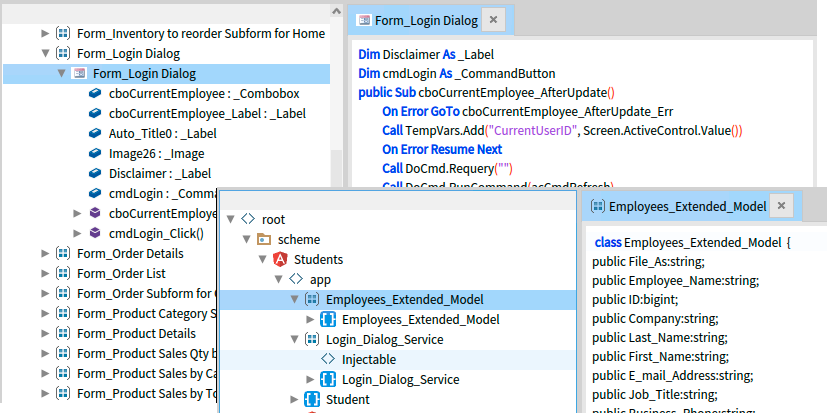}
		      \caption{Microsoft Access model (top left) and Typescript Angular model (bottom right) presented to the user when watching the source code}
		      \label{fig:srcTargetModels}
	      \end{figure}

	\item The user navigates the source hierarchy and selects an artefact to migrate.
		It could be a module, a form\footnote{In Access, forms are first-class citizens of the programming language}, a function,\ldots

	\item The user manually chooses where in the target hierarchy he wants to migrate the artefact.
		The tool provides migration actions for the user to perform.
		These actions are named \emph{directives} and introduced in \autoref{sec:directives}.
		The tool accepts any location; however, to be meaningful, it must be coherent with the artefact to migrate, \ie a function can be migrated into a class and become a method.

	\item The tool responds to the migration actions by applying migration rules (transformations) to perform the migration.
		This mechanism is explained in \autoref{sec:produce-directive-application} and \autoref{sec:map-directive-application}.

	\item The user has immediate feedback.
		If the tool can apply the directive properly, the user can immediately analyse and verify the code in the targeted model.
		If the tool fails, the user is informed, and the tool rolls back any intermediate change done by the failing directive.

	\item If the user is unsatisfied with the result after verifying the code, he can refuse it by rolling back to the previous state.
		Whatever the reason for a rollback (error or user choice), the migration expert (typically not the application developers) will need to add or change a rule for the process to restart.

	\item If the change is accepted, it is introduced in the target model.
		The user may continue applying directives, going back to step 2 or generating the code.

	\item The user chooses to apply all the modifications to the project's source code.
		This modifies the project files to comply with the new model.
		After this step, we restart the whole process from step 1.

\end{enumerate}

The tool allows executing the model modifications over the different application's source code at any moment, applying manual changes over any of the applications, and reloading the models.

\section{Elements of our migration infrastructure}
\label{sec:directive-rules}

Our approach enables developers to execute arbitrary partial migrations between two different application models  (see Section \ref{sec:application-model}) through the interactive application of  \emph{directives} (see Section~\ref{sec:directives}), which are resolved by the immediate or delayed application of rules (see Section \ref{sec:rules}).

This approach is an extension of the one presented by \etAl{Bragagnolo}{Brag22b}, enabling interaction and migrations beyond language migration.

\subsection{Modeling source and target applications}

\label{sec:application-model}
	Our model-driven approach applies modifications to target models based on entities and knowledge extracted from source models.
	All models are instantiated from the same heterogeneous unified meta-model.
	
	To achieve that, we load a model for each application in the migration project.
	In our primary study case, we target a client/server architecture. 
	We instantiate three different models, one per application:
	(i) legacy monolithic application (Microsoft Access), 
	(ii) migrated back-end application (Java Springboot) and 
	(iii) migrated front-end application (Typescript Angular).
	
\subsubsection{The Heterogenous Unified Meta-model}

	\label{sec:humm}
	\begin{figure}[htbp]
	\centering
	\includegraphics[scale=0.55]{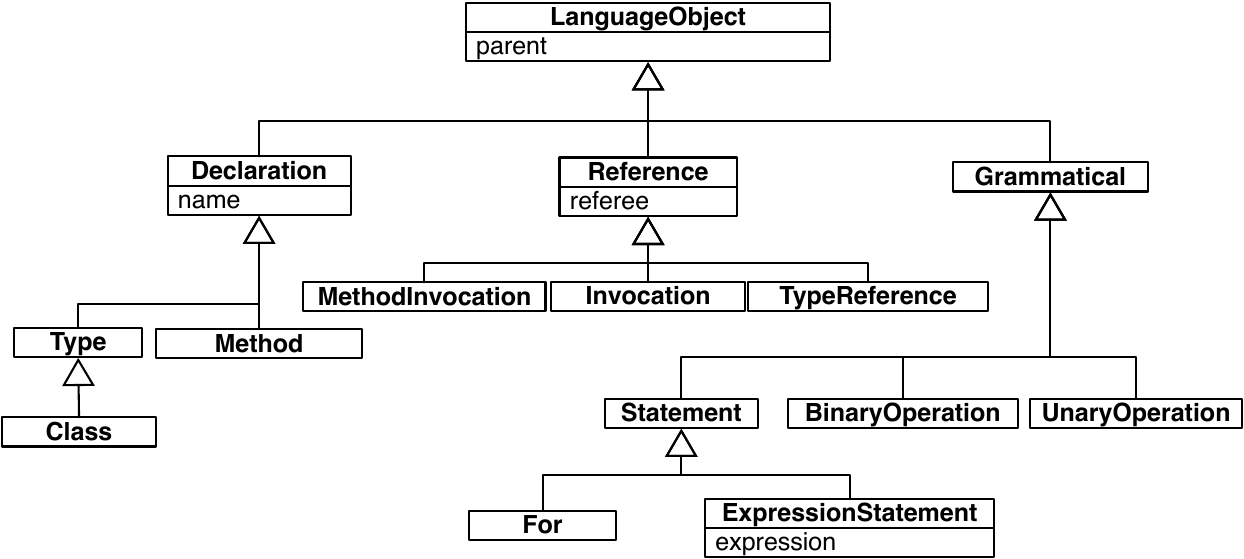}
	\caption{\label{fig:migration-meta-model} 
	Exempt of our Heterogeneous Unified Meta-Model.
	The three main entities in our model are Declaration, Reference and Grammatical. (See Text)
	}
\end{figure}

	We \emph{distinguish} different categories of types. 
	\autoref{fig:migration-meta-model} shows a fr
	Those that can be referred to (\eg variable declaration), that we call \emph{declaration nodes}; and nodes for elements of the language (like statements or arithmetic expressions) that we call \emph{grammatical}.
	On top of that, references are reified, giving a third kind: the \emph{reference}.
 	There is an implicit dependency between the nature of a declared artefact and the kind of reference able to use this artefact.
	To activate a function which receives one parameter, we use a function invocation with one argument.

    	\paragraph{Abstract Semantic Graph}
	Our approach leverages the heterogeneous unified meta-model presented by \etAl{Bragagnolo}{Brag22b}.
	This meta-model consists of an Abstract Semantic Graph (ASG) \footnote{An ASG is an Abstract Syntax Tree extended by linking uses with definitions or declarations, making it into a graph} that uses the same entities for different languages as long as the instances do have the same semantics: the modelling rationale behind is to model likeness and unlikeness of the different constructs.
	When this is not the case (\eg Java classes and TypeScript classes), we represent them with different types to be able to differentiate them in the migration rules.

	\paragraph{Representing \emph{inconsistent} intermediate states} 
	Our approach also requires representing \emph{inconsistent} intermediate states as it is iterative and interactive. 
	The migration engineer must be able to 
	(i) write a migration rule and see the outcome, whether it is or is not correct. 
	(ii) productive rules that produce partially correct outcomes due to a lack of information. 
	 \eg Let us consider that we have to migrate a function invocation. 
	 Suppose we do not know how the invoked function will be migrated. In that case, we cannot decide how to translate the invocation: method invocation, static method invocation, function invocation, or attribute access.

	\paragraph{Representing \emph{Primitive} and \emph{third party types} and \emph{behaviours}}
	Modelling of an application in a given programming language also includes creating declaration instances for the programming language's primitive types and any external library functions used. These declarations assert the existence of the library entities and have no complete definition; they are used to be referred by the parts of the application that require them.


\subsection{Directives: User explicit actions}
\label{sec:directives}

Our engine exposes two main actions to the user: \emph{Produce} and \emph{Map}.
These actions establish a relationship between two entities belonging to two different models. 
Within the duration of the action, one entity plays the role of the source, and the other plays the role of the target. 
To convey the direction of the action, we call them \emph{Directives}.

\paragraph{Directives}
\begin{description}
	\item[Produce:] Given a \emph{source entity} and a \textit{target context}, this directive instructs to produce the given \textit{target entity} inside the given \textit{target context}, based on the \emph{source entity}.  
	\item[Map:] Given a \textit{source declaration}, a \textit{target declaration}, and a \textit{scope of validity} (in the target model), this directive establishes a semantic equivalence between the two declarations, meaning that ---within the given scope--- past and future references to the source declaration will be replaced by references to the target declaration.
\end{description}
Directives are implemented by migration rules and mappings, as shown in the following sections.
These two directives offer the user the activation of two migrating actions.

\paragraph{Actions} 
		
\begin{enumerate}
\item  the \emph{declaration} of a software entity in a model (\eg a function in the legacy application) must be created as the \emph{declaration} of another software artefact in another model (\eg a method in the migrated application);
\item \emph{references} to a software artefact (\eg a primitive type or a library function in the legacy application) must be replaced by \emph{references} to another equivalent software artefact (\eg a matching primitive type or a replacing library function in the migrated application).
\end{enumerate}

Note that the first action (re-declaration of a software artefact) also entails the second, as the previous references to the original artefact must be changed to references to the migrated one.

\subsection{Cross-model Mappings \& Stubs}
		\label{ss:mapping}
		
		\paragraph{Mappings}
		Mappings are \emph{scoped} objects representing a \emph{semantic equivalence} between two models: 
		a source entity is equivalent to a target entity. \eg (long \arr bigint); (MsgBox \arr alert); etc. 
		
		By \emph{scoped}, we mean that they have a scope of validity.
		This is to say that A is equivalent to B for a specific project, package, class, method, etc. This context is defined when recording these mappings. 
		When the mapping is recorded automatically, this scope is the whole project. 
		When the mapping is recorded manually, the user defines its scope. 
		
		
		Some rules use mappings (See \autoref{sec:adaptivesrules}).
		In our approach, a mapping association is a result of a map directive (explained in \autoref{ss:mapping}) 
		or the execution of a Productive Rule, as explained in \autoref{sec:producerules}.
		\paragraph{Stubs}
		A reference node can only point to a declaration node within the same model.
When a reference in a model (\eg a migrated application model) should point to a declaration that does not exist in this model (typically a declaration that has not been migrated yet), we create a \emph{stub declaration} in the same model as the reference.
The reference may then points to this stub, and the stub itself points to the declaration of the entity in the other model.
Therefore \emph{stubs} are bridges between models.
As long as it contains stubs, a model cannot generate valid source code because some references still need to be migrated.
A specific mechanism replaces these stubs with actual declaration nodes when the latter are created (Section~\ref{sec:produce-directive-application}).

\subsection{Rules}
\label{sec:rules}
	
	Rules are \emph{scoped} conditional operations consisting of a Condition and an Operation. 
	\emph{Condition} consists of a predicate that ensures the satisfaction of the operation's assumptions and requirements.
	\emph{Operation} consists of any systematic modification over the target. 	
	By \emph{scoped}, we mean that they have a scope of validity. 
	This scope is all migration models, as there are default rules, a specific project, package, class, or method. 
	This context is defined when installing the rules.	
	We have two families or rules: Productive and Adaptive. 
	\subsubsection{Productive Rules}
	\label{sec:producerules}
	These rules create a migrated version of a source entity in the context of a target entity, that is to say, as a child of this target entity.
	Applying a Productive Rule also entails automatically mapping the source entity and the produced target entity. 
	
	\emph{Condition} consists of a predicate that tells if the rule can produce this migrated version within the specific target entity. 
	
	\emph{Operation} consists of modifying the target entity by defining the source entity's migrated version. 
	These rules are typically applied from the legacy application model ---containing the source entity--- to a migrated application model ---containing the target entity.
	
	For example, a rule that is executed when the source entity is a function and the target context is a class, and it operates a migration by defining a method in the target class based on the source function.

	\subsubsection{Adaptive rules}
	\label{sec:adaptivesrules}
	We said before that \emph{references} to a software artefact must be replaced as \emph{references} to another equivalent software artefact.
	To enable this replacement mechanism, we copy the \emph{references} as they are, knowing that this may be a wrong decision,
	 and delay the migration criteria to when this other software artefact is available.
	Adaptive rules detect the apparition of semantically equivalent declarations and modify or replace reference objects based on the nature of the software artefact. 
	
	\emph{Condition} consists of a predicate that tells if the rule can modify or replace a target reference to refer to a given target declaration.
	
	\emph{Operation} consists of the adaptation of the reference object. 
	
	For example, let us consider a function that just migrated as a static method.
	 Right after, an adaptive rule will detect that this static method is recognised as equivalent to the function and therefore detect all the uses (migrated before and after the migration of the function) and adapt them from the form \texttt{function()} to the form \texttt{ClassName.function()}.

\subsection{Migration engine: Executing directives with rules look-up and default behaviour}
\label{sec:migration-engine}

At the heart of our migrating approach is the \emph{engine} controlling the application of the directives, the selection of the rules and their application.
In our approach, rules and mappings are contextualised, meaning they only affect a part.  
We can install a rule or a mapping; thus, it is always applicable or only applicable in the context of specific construction (project, package, class, etc.).

\subsubsection{Executing the produce directive}
\label{sec:migration-produce}
When the user initiates a produce directive on a source entity and a target context, the engine searches in the target context for a \emph{Productive Rule} that can handle the migrating case. This is to say that the condition predicate accepts the source entity and target context. 
As we explained in \autoref{sec:rules}, rules are contextualised, so we look up rules by context. 
The search starts in the target context and goes up the ASG as needed. 
We present in detail its application in \autoref{sec:produce-directive-application}.

The produce directive is direct order of transformation from the user. 
Therefore, it must be accomplished immediately, which means we must have a default rule.

\paragraph{Default Productive Rule: AnyCopy}

As we explained in \autoref{sec:humm} we base our approach on a heterogenous unified meta-model, meaning that we use the same meta-model for all projects. 
This choice is no accident. We want only to provide rules for the cases that require a transformation. 
For the rest, we provide the AnyCopy Productive Rule. 
		\paragraph*{AnyCopy}
		\begin{description}
			\item[Context:] Root context (available regardless of which project is the target)
			\item[Condition:] ~
				\begin{enumerate} 
				     \item Allways true
				\end{enumerate}
			\item[Operation:] ~
				\begin{enumerate} 
					\item Make an instance of the same class as the source entity within the target context.
					\item Migrate each of the children of the source entity using the freshly created instance as the target context.
				\end{enumerate}
		\end{description}

\subsubsection{Executing the Map directive}
	When a map directive is initiated by the user, on a source and target declarations, within a context, the engine installs a mapping in the given target context (for example, function F is equivalent to method M only in the package P) right after; the engine searches all the references within the given context to be fixed.
	 Each of them searches for any \emph{adaptive rules} that can be applied after the mapping has been registered, starting with the reference-to-fix context and going up the ASG as needed.  
	 Adaptive rules are executed only when there is a mapping available. Therefore there is no default rule. 
We present in detail its application in \autoref{sec:map-directive-application}.

\section{The approach in action}
\label{sec:approach-in-action}
We illustrate the process of the migration of the VBA sub-procedure ``showName'' (Listing~\ref{lst:showName-sub}), to the context of a Java class named ``MyDestination'' in the back end application (Listing~\ref{lst:javaTarget}).
The source sub-procedure is a simple piece of Visual Basic that pops up a dialogue showing the content of a variable \ascode{name}, concatenated with the string literal ``Ms ''.
Because this is migrated to the back-end application, the migrated class cannot access a GUI.
Any information that was previously displayed should now be logged.

\begin{lstlisting}[float=htbp, language=VBScript, caption=VBA sub-procedure that pops up a dialog, label=lst:showName-sub]
Dim name as String
public Sub showName()
	Call MsgBox ("Ms " & name)
End Sub
\end{lstlisting}

\begin{lstlisting}[float=htbp, language=Java, caption=MyService Java class, label=lst:javaTarget]
	package MyPackage; 
	class MyDestination {
		public static void log (String) { 
		 ...
		}
	}
\end{lstlisting}

\subsection{Engine setup} 

We consider the engine configured with specific rules and no previously existing mapping for this example. 
We now present the installed productive rules AnyCopy, CopyAsStaticMethod and CopyReplaceOperator and the adaptive rule RenameAdaptToStaticReceiver.
\subsubsection{Productive Rules}
\label{sec:produce-example-rules}
\paragraph*{AnyCopy} was presented in \autoref{sec:migration-produce}. 

\paragraph*{CopyAsStaticMethod}
	\begin{description}
			\item[Context:] Java migrated application 
 			\item[Condition:] ~
				\begin{enumerate} 
				     \item The source entity is a sub-procedure \textbf{AND}
				      \item The target context is a class
				\end{enumerate}
			\item[Operation:] ~
				\begin{enumerate} 
					\item Define in the target entity a static method with the same selector as the source entity and using void as returning type reference.
					\item Migrate all the children of the source entity using the method as the target context. 
				\end{enumerate}
		\end{description}

\paragraph*{CopyReplaceOperator}
	This rule requires two parameters when it is instantiated: The operator to detect (OtD) and the operator to replace it (OtR). 
	\begin{description}
			\item[Context:] Java migrated application 
 			\item[Condition:] ~
				\begin{enumerate} 
				     \item The source entity is a binary operation
				     \item The operator matches OtD.
				\end{enumerate}
			\item[Operation:] ~
				\begin{enumerate} 
					\item Define a binary operation in the target entity, using OtR as an operator. 
					\item Migrate all the children of the source entity using the binary operation as the target context. 
				\end{enumerate}
		\end{description}

\subsubsection{Adaptive Rule} 
\label{sec:adaptiverules-static}
For this example, we use only one adaptive rule, defined before by \etAl{Bragagnolo}{Brag22b}.

\paragraph*{RenameAdaptToStaticReceiver -- Replace by method invocation with \emph{static} receiver }
	\begin{description}
			\item[Context:] Java migrated application 
 			\item[Condition:] ~
				\begin{enumerate} 
				      \item The target reference is a function invocation \textbf{AND}
				      \item  The given map target declaration is a method \textbf{AND}
				       \item The given map target declaration is \emph{static}.
				\end{enumerate}
			\item[Operation:] ~
				\begin{enumerate} 
					\item Define a method invocation expression (a reference) using the method's parent as a receiver. 
					\item Set the arguments used by the function invocation in the new method invocation.
					\item Set the new target reference to the mapped target declaration. 
				\end{enumerate}
		\end{description}

\subsubsection{Rule installation} 
\begin{figure}[htbp]
\begin{center}
\includegraphics[width=0.8\textwidth]{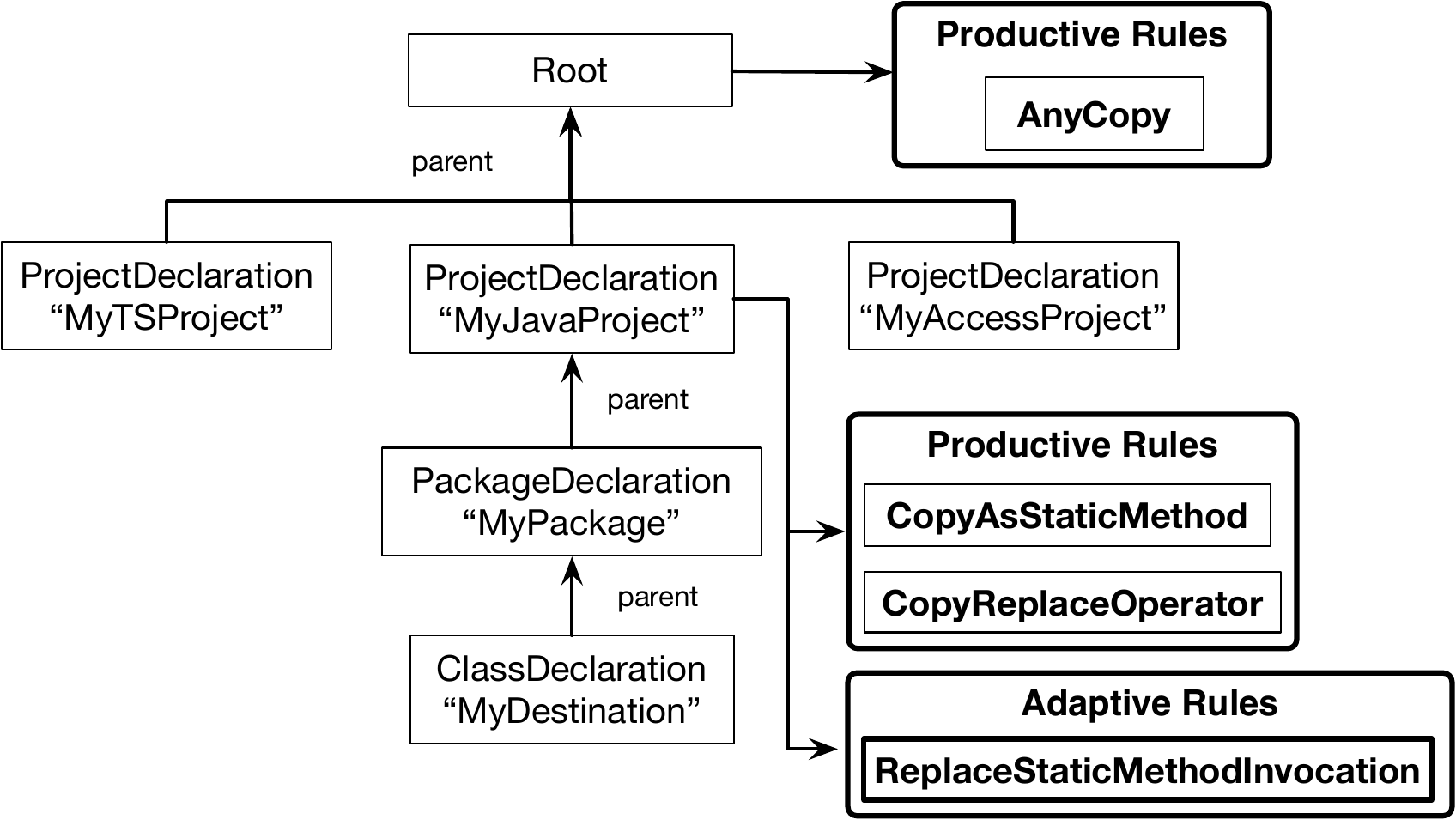} 
\caption{Rules by target context. Between the contexts, we find all the projects. 
We only analyse the rules related to the Java project. 
At root, we find the rules that apply to all targets. 
In the Java level we find Productive and Adaptive rules.}
\label{fig:ruleinstall}
\end{center}
\end{figure}

To complete the engine setup, we must install the rules we want for our example. 
As we said in \autoref{sec:rules}, rules are \emph{contextualized}. As we said in \autoref{sec:migration-engine} the engine performs a lookup through the target to find rules and mappings.  
\autoref{fig:ruleinstall} depicts our example installation. 
AnyCopy is always available regardless of the targeted project.
In the context of the target project, we find the rules defined above: CopyAsStaticMethod, CopyReplaceOperator and RenameAdaptToStaticReceiver. 

\subsection{Example of the application of a produce directive}
\label{sec:produce-directive-application}

The produce directive starts when a user asks to migrate the specific source declaration, the sub-procedure ``showName'', into a specific target declaration, ``MyDestination'' class.

\begin{sidewaysfigure}[htbp]
\begin{center}
\includegraphics[width=\textwidth]{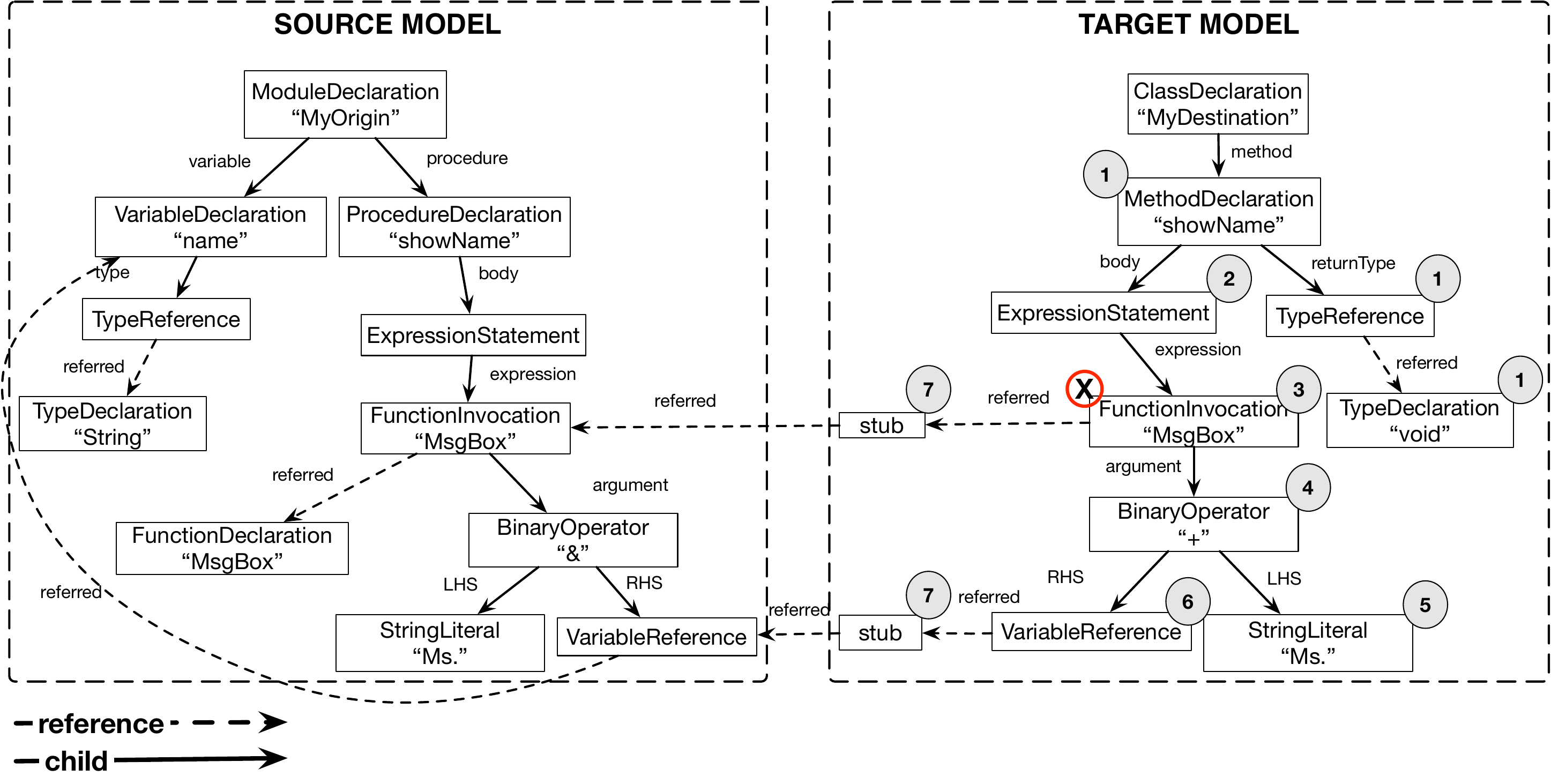} 
\caption{Produced ``showName'' method from ``showName'' sub-procedure. BinaryOperator and StringLiteral are grammatical nodes; the FunctionInvocation node is not valid in the target model. The numbers in the visualisation refer to the step when the different elements have been created. We note the difference between children and references.}
\label{fig:showNameAST}
\end{center}
\end{sidewaysfigure}

The result of the application of the produce directive is displayed in Figure~\ref{fig:showNameAST} and can be used to follow the migration.
We break down the example into eight steps. 
\autoref{fig:showNameAST} shows the source model and the target model after the execution of this directive. 
Each number attached to a target model entity is the step where they are created. 

\begin{description}
	\item[Step 1] The engine searches for a \emph{Productive Rule} in the context of the ``MyDestination'' class (of the target model) and does not find it; next, it looks in its parent context and up in the context hierarchy until the top level node of the Java migrated application model where the \rulename{CopyAsStaticMethod} rule is found that matches: following its definition (see \autoref{sec:produce-example-rules}), 
	it accepts a subprocedure being migrated to a class.
	The rule creates a method declaration node as a child of the class declaration node (the target context) and gives the name of the sub-procedure to the method. 
	In this case, the return type is set to \ascode{void} as VBA sub-procedures do not return any value. 
	The method is also defined as \ascode{static} to allow invoking it without an object. 
	Then the rule delegates the migration of all children to the engine. 
	\item[Step 2]  Here, ``showName'' has no parameter, so the only child is the single statement in the sub-procedure's body.
The engine now looks for a \emph{Productive Rule} accepting the statement as source entity and the newly created method declaration node as target context. The rule matching these source and target entities is found at the top level of the model. It is the generic \rulename{AnyCopy} rule (Defined in \autoref{sec:migration-produce}) that copies the source sub-tree into the destination context.
So it will copy the ``expression-statement'' node (a grammatical node, see Section~\ref{sec:application-model}) of the sub-procedure.
And again, delegate handling of its child (a ``function call'' reference node).
	\item[Step 3]
This also falls back to the \rulename{AnyCopy} rule.
But this time, it must be noted that the copied node is not valid in a Java model (no function calls in Java, only method calls).
This problem will be corrected later (by a map directive).
For now, the reference remains a function call and points to no declaration node.
	\item[Step 4]
	The next step is to handle the children of the function call. In this case, it is the argument of the function: the binary operation.
	This time \rulename{CopyReplaceOperator} (defined in \autoref{sec:produce-example-rules}) rule is found that matches.
	This rule has been configured to detect and replace the ``\&'' binary operator (string concatenation), which is lookup in the top context of the Java model that generates a ``+'' operator.
	\item[Step 5 and 6]
	Two new iterations of this process will apply the \rulename{AnyCopy} rule on the operator's arguments.
As the first argument, this gives a literal string node (valid in Java) and, as the second argument, a reference to a variable named ``name''.
Note that at this stage, the reference node points to nothing (no declaration node).
	\item[Step 7]
	The engine is now done with the creation of nodes in the target model, the entire AST of the ``showName'' sub-procedure declaration has been recreated (in a modified form) in the Java model, and there are two references (variable access and function call) pointing to nothing.

The engine then looks at the references that were created.
For the variable reference, it searches whether an adaptive rule can make this variable reference fit the target model (it should be pointing to an existing variable).
Supposing there is none (the variable still needs to be migrated), the engine creates a stub declaration to which the variable reference can point, which points to the variable reference in the source model.
Similarly, for the function call, the engine looks for an existing adaptive rule able to adapt the call to \ascode{MsgBox} to fit the target model (it should be pointing to an existing method) and does not find one.
A stub function declaration is created that points to the function invocation in the source model.
The state of the models at this stage is pictured in Figure \ref{fig:showNameAST}.
	\item[Step 8]
	Finally, in the context associated to``MyDestination'' target class, a mapping is registered from the legacy ``showName'' function to the migrated ``showName'' method.
At this stage, the produce directive is finished; it gives a target model that is not coherent as it contains nodes that are not valid in Java.  Invalid nodes are allowed temporarily if one does not try to export the model to the source code.

\end{description}

Please, note that two features of our meta-model are leveraged in this explanation. 
(i) Four out of six rule applications use the \rulename{AnyCopy} rule.
This is due to our application meta-model rationale based on similarities and differences.
(ii) The application of the directive finished in an expected yet incorrect state. 
This is possible thanks to the model's ability to represent \emph{inconsistent} intermediate states.

Such is the interest in using a heterogenous unified meta-model, able to copy as default behaviour and represent inconsistent intermediate states.

\subsection{Example of the application of a map directive}
\label{sec:map-directive-application}

Having converted the VBA ``showName'' sub-procedure to a Java method, the user is notified that this method is incomplete: it requires the resolution of two artefacts in the target model: (i) the invoked method ``MsgBox'' (ii) the referred variable ``name''.

For helping to resolve these artefacts, our approach includes an action to manually express that these missing artefacts have an equivalent in the target model: the map directive. 


To do this, the user applies a map directive (see Section~\ref{sec:directives}) that is going to register a mapping between the source declaration of ``MsgBox'' (a \emph{declaration} that contains no definition as it is a VBA library routine), and the target declaration of Java method ``log'', within the context of ``MyDestination'' class. 
Each time a mapping is registered, either by the map directive or as a result of migration a declaration (step 8 of produce directive \autoref{sec:produce-directive-application}), the engine starts what we call an \emph{adaptive phase} (systematic application of adaptive rules). 

\begin{figure}[htbp]
\begin{center}
\includegraphics[width=0.9\textwidth]{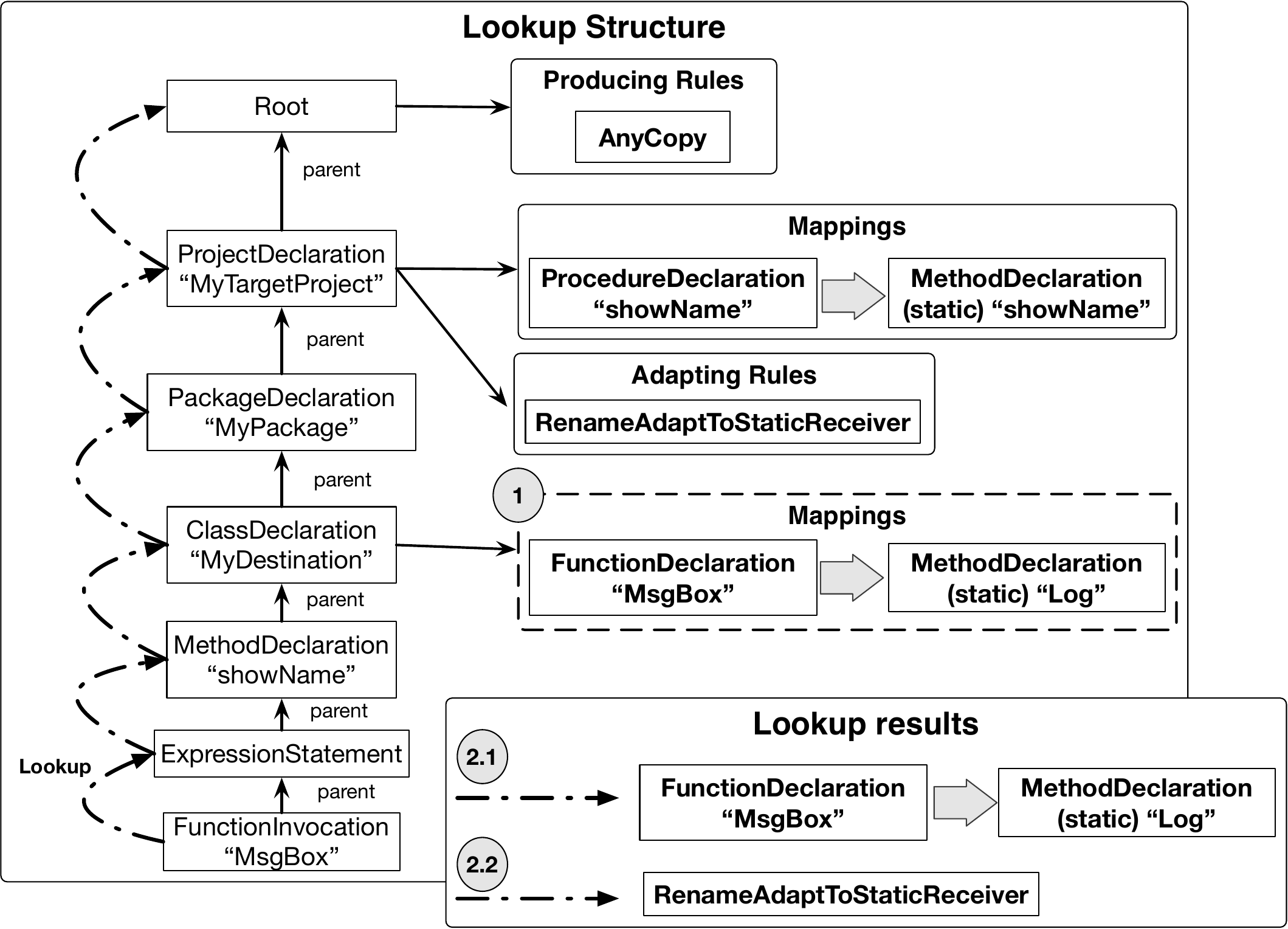} 
\caption{Map directive application: Mapping registering, mappings lookup, adaptive rule lookup. The mapping (ProcedureDeclaration ``showName'' \arr MethodDeclaration ``showName'') was automatically registered during the Produce directive
The numbers in the visualisation refer to the step related. Step 1 creates a mapping. 
Step 2 does a double lookup selecting specific items. }
\label{fig:map-directive-application}
\end{center}
\end{figure}

We break down the map directive into four steps, including registering the mapping and the \emph{adaptive phase}; the first step and the two sub-steps 2.1 and 2.2 are illustrated in \autoref{fig:map-directive-application}.

The application of adaptive rules is made with a kind of ``double lookup''.

\begin{description}
	\item[Step 1] The engine registers the map between the source and target declarations in the list of mappings of the application context. (MsgBox \arr MyDestination.log) 
	\item[Step 2] First step of the adaptive phase. The engine finds all unresolved uses of the source entity in the target model: all the reference objects pointing to stubs pointing to this source entity. 
	\begin{description}
		\item[Step 2.1] 
		The first lookup aggregates all the possible mappings affecting each of the references. As shown by \autoref{fig:map-directive-application}, this lookup starts from the context of each reference. This aggregated information is an ordered list. The most concrete mapping is tested first. In our example, the only mapping responding to our case is the one just installed.
		\item[Step 2.2] 
		The second lookup tests all possible adaptive rules with each reference and each mapping in order. 
		As shown by \autoref{fig:map-directive-application}, this lookup starts from the context of each reference.
		This process yields the first rule that tested positive for its application.
		If no rule is found, the process finishes.
		In our example, the first (and only) rule testing positive is \rulename{RenameAdaptToStaticReceiver}.
	\end{description}
	\item[Step 3] The process finishes here if no rule is found. If a rule is found, the engine applies it. 
	In our example, the engine applies the \rulename{RenameAdaptToStaticReceiver} parametrised with the function call and the mapping that represents equivalence between ``MsgBox'' and ``MyDestination.log''. 
	As an outcome, the rule replaces the function call with a method invocation, using the same arguments as the original call, and setting \texttt{MyDestination.log} method as referee. 
	\item[Step 4] Last step of the adaptive phase. The engine removes all the useless stubs. 
\end{description}
 
In this way, the mechanism can be applied retroactively (by searching for past Stub nodes) and to future migrations (because the produce directive registers mappings for all migrated declarations).

\section{Validation }
\label{sec:valid}

We implemented the engine and interactive UI to validate our approach and conducted several experiments. 
In this section, we overview the implementation and present different results.

\subsection{Implementation}
\label{sec:implementation}
\label{sec:impl}
To validate the approach, we implemented the heterogeneous unified meta-model ASG, the migration engine and the UI tool allowing the user to express the migrating directives.

\begin{figure}[htbp]
	\centering
	\includegraphics[scale=0.20]{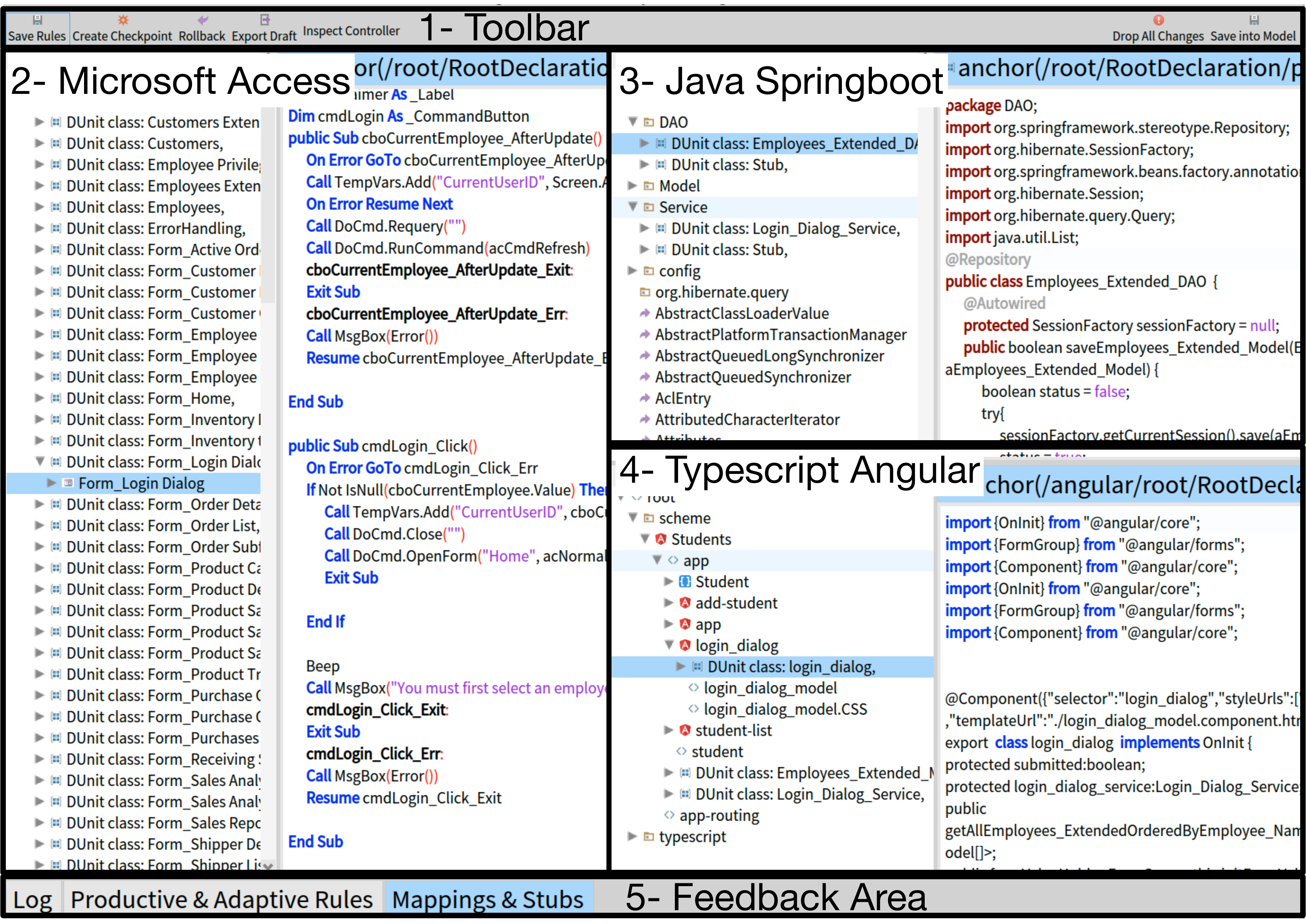}
	\caption{\label{fig:screenshot} Tool Screenshot. The tool consists of a toolbar, three areas of model visualisation: one for MS Access, one for Java and one for Typescript, and an area for feedback}
\end{figure}

\begin{figure}[htbp]
	\centering
	\includegraphics[scale=0.30]{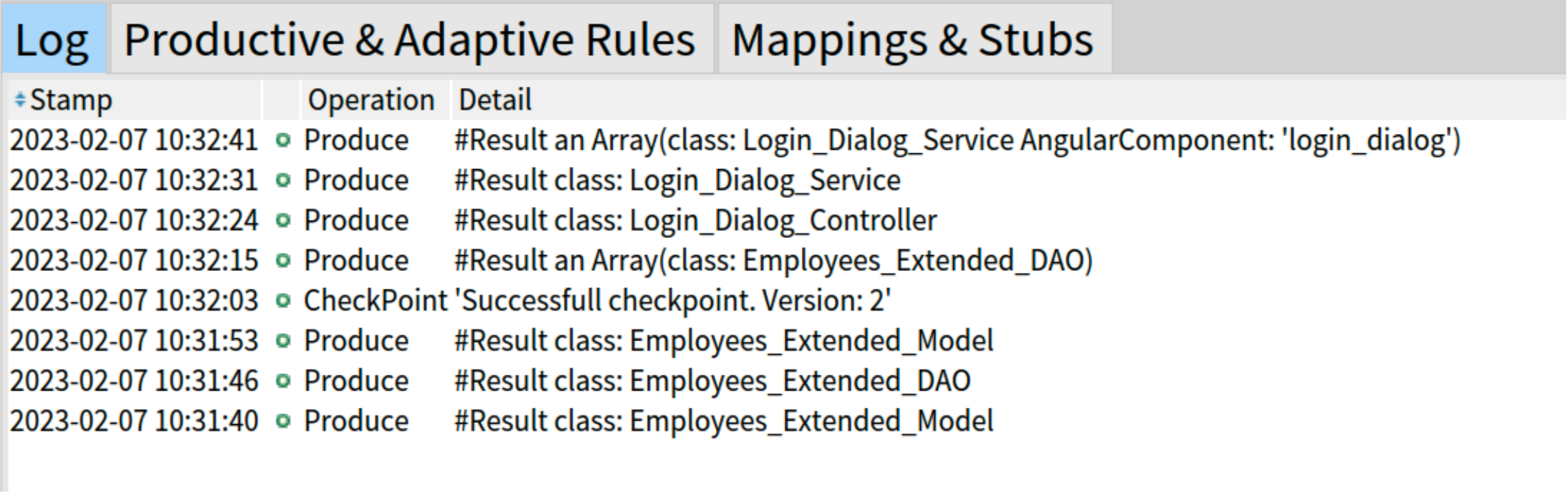}
	\caption{\label{fig:screenshot-log} Tool feedback: The log panel. Here we see the outcome of the different user interactions.}
\end{figure}

\begin{figure}[htbp]
	\centering
	\includegraphics[scale=0.27]{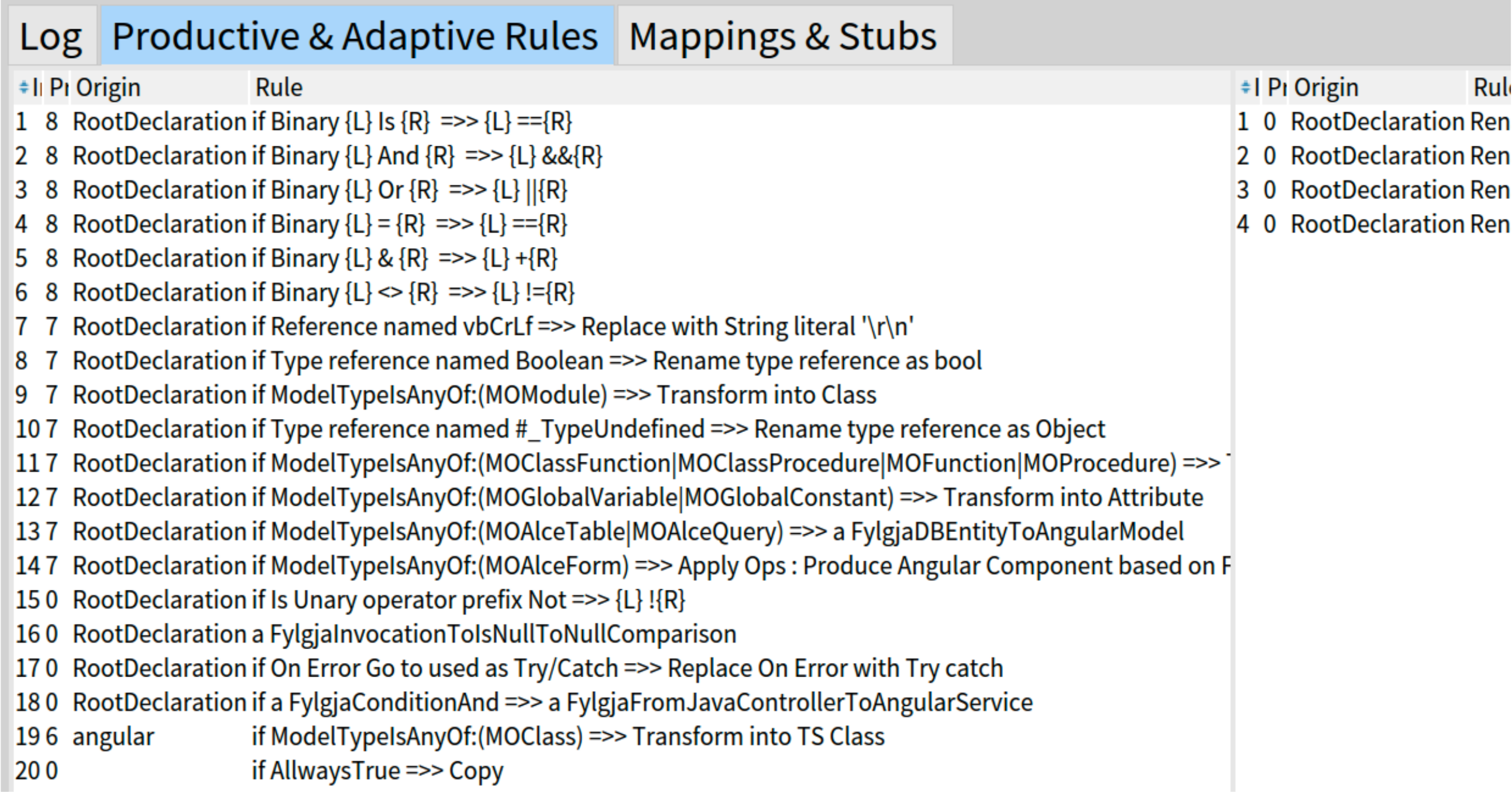}
	\caption{\label{fig:screenshot-rules} Tool feedback: The rules panel. On the left panel, we observe the productive rules. On the right one, the adaptive rules}
\end{figure}

\begin{figure}[htbp]
	\centering
	\includegraphics[scale=0.27]{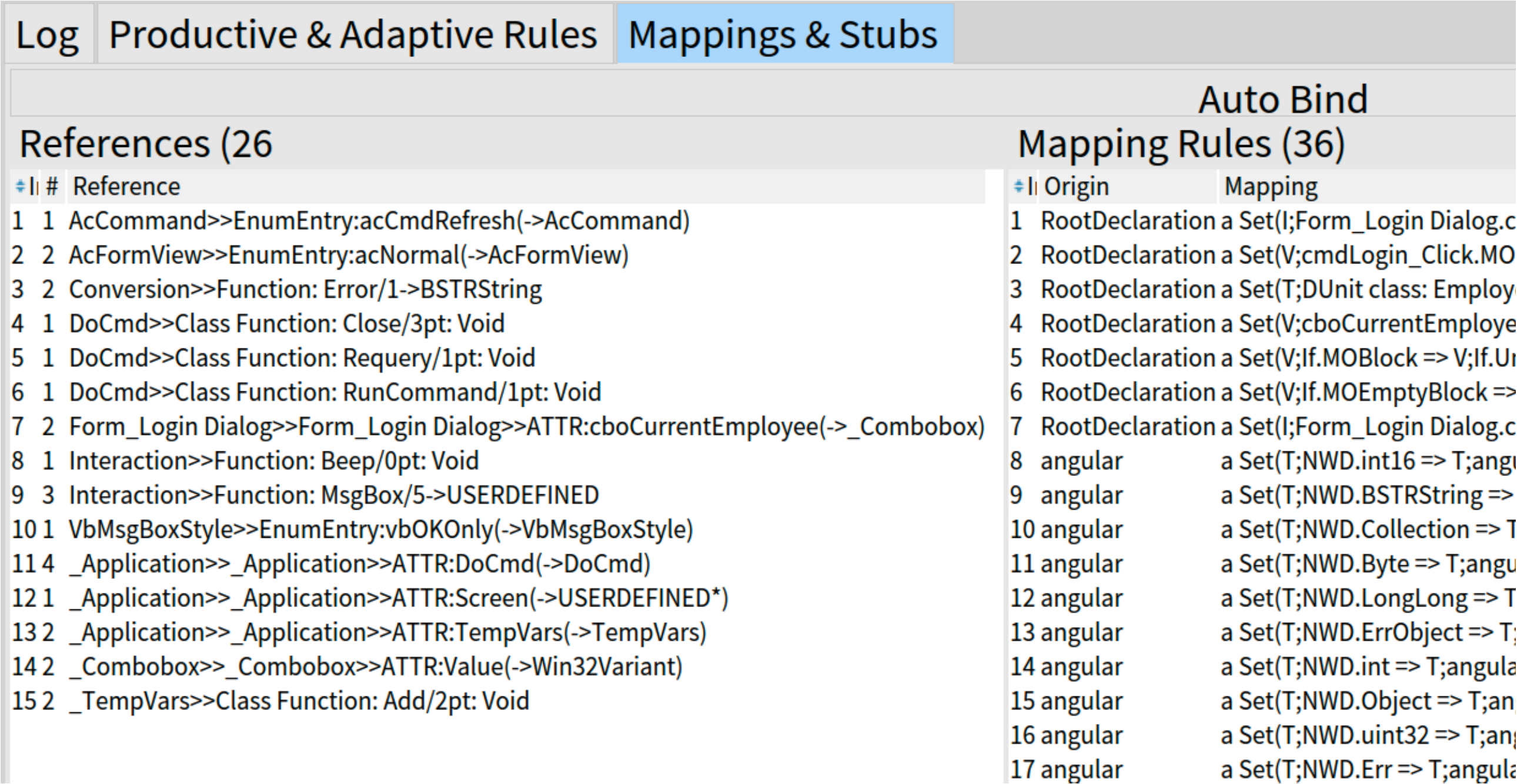}
	\caption{\label{fig:screenshot-stub} Tool feedback: The stubs and mappings panel. On the left part, we see the Reference table, which shows the elements referred to by the stubs. On the right panel, are the established mappings.}
\end{figure}

On the one hand, \autoref{fig:screenshot} depicts the central area of the migrating tool.
On the other hand, \autoref{fig:screenshot-log}, \autoref{fig:screenshot-rules}, and \autoref{fig:screenshot-stub}  depict the three feedback areas of our implementation.


\begin{itemize}
\item[1-] Toolbar: tools for the project.
\item[2-] The left panel shows the structure and code of the MS Access project. 
\item[3-] The right-top panel shows the structure and code of the Java-SpringBoot project.
\item[4-] The right-top panel shows the structure and code of the TypeScript-Angular project.
\item[5-]{Feedback Area} \autoref{fig:screenshot-log} shows logs of the actions applied, informing errors and successes. \autoref{fig:screenshot-rules} shows the rules valid in a selected context (a class by example). \autoref{fig:screenshot-stub} shows the mappings and stubs tracked so far and valid in a given context. 
\end {itemize}

In the feedback area, we can find three tabs.
In the log tab, we find all the successful and failed interactions.

The production tab is bound to the last selected declaration within any model. It shows all the applicable productive rules available from this declaration context.

The mappings \& references tab is also related to the last selected declaration within any model.
The right side shows the applicable adaptive rules available from this declaration context.
The left side shows all the references that are not resolved within this declaration.

\paragraph{Using the tool.}
The user can drag and drop a Declaration from one model to another.
Once the drop is done, a popup asks the user to tell which directive she wants to apply: map or produce.
In implementing the approach, we decided to allow the user to have different levels of control over the lookup of rules when applying the produce directive.
(i) automatic lookup: the lookup uses the first found result. 
(ii) multiple choice: the lookup prompts the user to choose the rule to be applied. 
(iii) debugging: the lookup prompts the user to choose each rule.

In both (ii) and (iii) cases, the user chooses from all the rules available in the scope with positive conditions.

\begin{figure}
	\centering
	\includegraphics[scale=0.25]{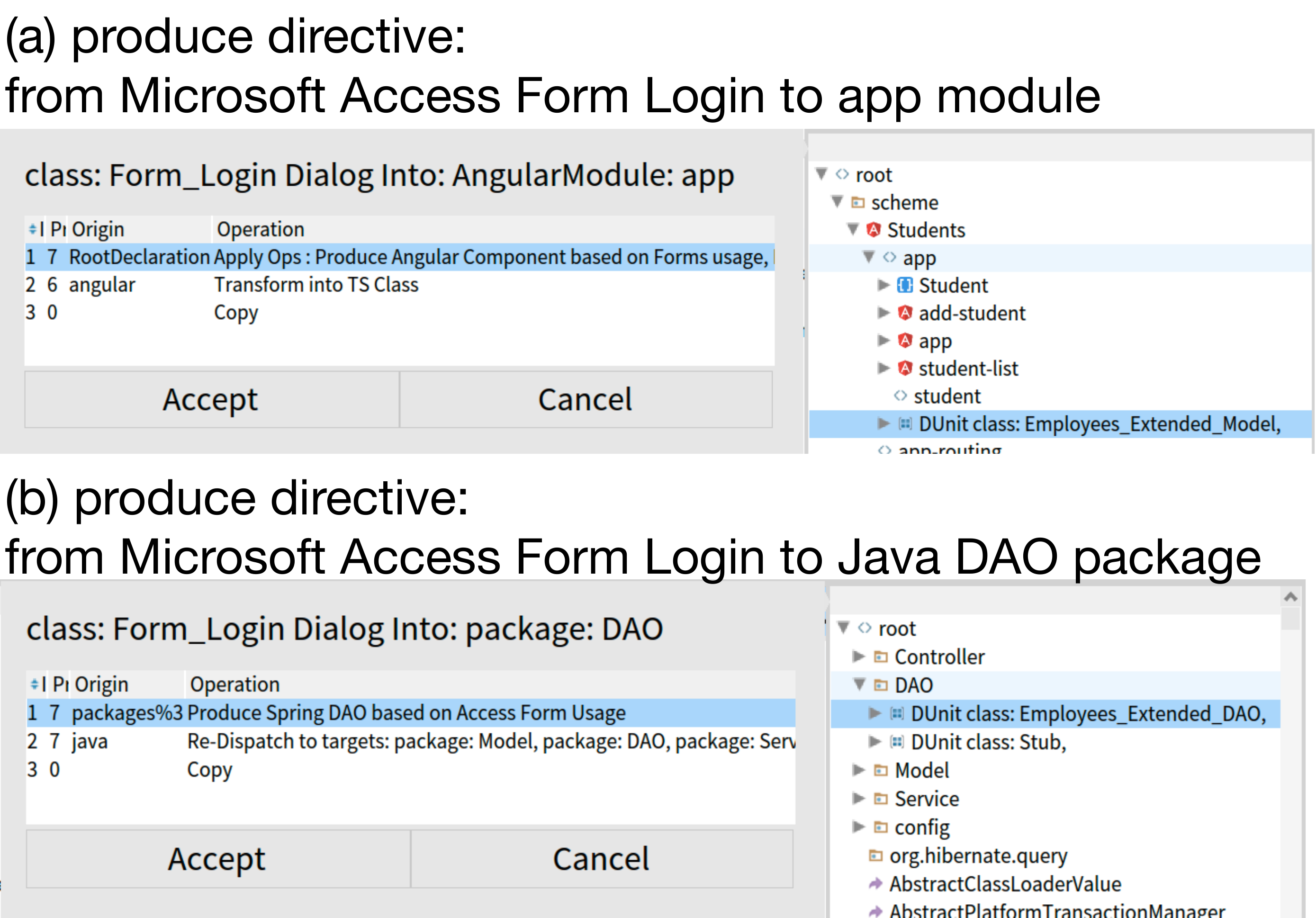}
	\caption{\label{fig:choose-rule} Choosing a rule. The figure shows the GUI prompting the user to choose the rule to apply. 
	The rules displayed in the table are all available in the scope with positive conditions. 
	We can see that regardless being the source the same entity, the rules in (a) are different to those in (b) as the target is different.}
\end{figure}

\autoref{fig:choose-rule}  shows the GUI prompting the user to choose the rule to be applied in two different cases:
\autoref{fig:choose-rule}(a) prompts for the rule when migrating the Form login to the ``app'' Angular module while 
\autoref{fig:choose-rule}(v) prompts for the rule when migrating it to the ``DAO'' Java package. 

\subsection{Validating experiments} 

\subsubsection{Library and paradigm partial migration} 
\label{sec:lib-migration}
Presented in \etAl{Bragagnolo}{Brag22b}. 
We produced 53 test unit tests in Microsoft Access using the 53 most used library artefacts in an industrial project, based on analysing and sorting the artefacts by the number of uses. 
All the function calls are automatically transformed into method invocation.  
We defined five productive rules and five adaptive rules. 
We performed two migrations of these tests: one to Java and one to Pharo. 

\paragraph{Rules} 
We defined four \emph{Productive} rules: AnyCopy, ModuleToClass, FunctionToMethod, GlobalToAttribute and CopyReplaceBinaryOperator. 
We defined five \emph{Adaptive} rules: SimpleRename, RenameAdaptToStaticReceiver, RenameAdaptToThisReceiver, RenameAdaptToArgumentReceiver and Autowrap. 
The autowrap rule generates the skeleton of whatever used element defined in a library that has no mapped equivalent in the target.
We installed \textbf{the same rules} for both targets, Pharo and Java. 
The only difference between the configurations was the source and target ASG mappings.

\paragraph{Mappings} 
The mappings differ by target since the available libraries are different. \autoref{tab:mappings} shows few of them. 
We found mappings for functions, types and constants.
Please address \etAl{Bragagnolo}{Brag22b} for the mapping details.
\begin{table}[htbp]
  \centering
   \caption{Type mappings Mapping examples table}
    \begin{tabular}{p{0.15\linewidth}  p{0.15\linewidth}   p{0.35\linewidth}}
    \textbf{VBA} & \textbf{Java} & \textbf{Pharo} \\
    \midrule
    VBA.Single & java.lang.Float & Kernel-Numbers.Float \\
    VBA.String & java.lang.String & Collections-Strings.String \\
    VBA.ubyte & java.lang.Byte & Kernel-Numbers.SmallInteger \\  
    \end{tabular}%
  \label{tab:mappings}%
\end{table}%

\paragraph{Results}

\begin{table}[htbp]
  \centering
  \caption{Translation Results. Mapped: number of library elements mapped. Parsed: number of results that are verified by a parser. Compiled: number of results that are successfully compiled in the target technology. Test Result: Number of compiled tests's results: Success, Failure or Error.  }
    \begin{tabular}{lrrrrrrr}
    Test case & \multicolumn{1}{l}{Total} & \multicolumn{1}{l}{Mapped} & \multicolumn{1}{l}{Parsed} & \multicolumn{1}{l}{Compiled} & \multicolumn{3}{c}{Test Result} \\
          &       &       &       &       & \multicolumn{1}{l}{~S} & \multicolumn{1}{l}{~F} & \multicolumn{1}{l}{~E} \\
          \midrule
	  Constants & 14    & 1     & 14    & 1     & 1     & 0     & 0 \\
           Functions & 21    & 11    & 21    & 16    & 8     & 3     & 5 \\
           Type & 18    & 13    & 18    & 9     & 4     & 4     & 1 \\
     \textbf{Java Total} & 53    & 25    & 53    & 26     & 13     & 7     & 6 \\
     \midrule
    Constants & 14    & 4     & 14    & 14    & 3     & 0     & 11 \\
    Functions & 21    & 14    & 21    & 21    & 10    & 3     & 8 \\
    Type  & 18    & 13    & 18    & 18    & 7     & 10    & 1 \\
    \textbf{Pharo Total} & 53& 31    & 53   & 53   &  20   &  13    &  20    \\
    \end{tabular}%
     \begin{tablenotes}
        \item {Test Legends:} (S)uccess, (F)ailure, (E)rror.
    \end{tablenotes}
  \label{tab:java-results}%
\end{table}%

\autoref{tab:java-results} gives the results per target. 
By column: \emph{Mapped:} the number of library elements mapped. 
\emph{Parsed:} the number of results verified by a parser. 
\emph{Compiled:} the number of results successfully compiled in the target technology. 
\emph{Test Result:} the number of compiled tests’ results: Success, Failure or Error.

We can see a relation between the number of mapped entities and the results of the amount of migrated tests that are success, failure or error.
For Java, we mapped 47\% of entities. 58\% of them in Pharo.
For both targets, the produced source code is verified by the SmaCC Java \footnote{https://github.com/j-brant/SmaCC} parser and the Pharo parser. 
With ten rules that are the same for two different languages, we produce 100\% of parseable code. 
The Java compiler compiles 47\% of the translated testing methods.
The Pharo compiler compiles 100\%. 
The compiling rate seems tightly related to the mapping in the case of Java since the compiling errors come from wrong typing.
We argue that most of them stem from the MS Access type named Variant, which we mapped to Object in Java. 
 MS Access Variant accepts any value but keeps the type of the element (object or not). 
In Java, the type Object accepts only objects (no primitive values allowed), but if a variable is typed as Object, it loses all singularity.

\subsubsection{Migrating Tables and Queries to a Java back end and a Typescript frontend}
\label{sec:table-migration}

We migrate all the tables and queries of the Microsoft Northwind project \footnote{ Microsoft Access Northwind traders, learning example }.
Northwind consists of 20 tables and 27 queries.

\paragraph{Rules} 
In this case, the \emph{Productive} rules differ for the Java back end and Typescript front end. 
\begin{description}
	\item[Back end] ~
	\begin{description} 
		\item[TableOrQueryToJavaPersistentModel] Defines a class with persistent mappings based on the given Table or Query. It adds public accessors.  
		\item[TableOrQueryToDAO] Defines a DAO which uses the class produced by the previous rule. It defines default methods for CRUD actions.
	\end{description}
	\item[Front end] ~
	\begin{description} 
		\item[TableOrQueryToClass] Defines an exported public class based on the given Table or Query. 
		All the attributes are public for accessing the content directly. 
	\end{description}
\end{description}
The \emph{Adaptive} rule is the same for both the front and back end: SimpleRename.  
This rule is in charge of adapting all the types based on mappings. 

\paragraph{Mappings}
The mappings for this experiment are presented in \autoref{tab:mappings-talbes}.
 \begin{table}[htbp]
  \centering
   \caption{Mapping examples table}
    \begin{tabular}{p{0.30\linewidth} p{0.22\linewidth}  p{0.22\linewidth}}
    \textbf{VBA} & \textbf{Java} & \textbf{Typescript} \\
    \midrule
    Access.dbText & java.lang.String &string \\
    Access.dbMemo & java.lang.String & string \\
    Access.dbDate & java.time.LocalDate & Date \\
    Access.dbAttachment & java.sql.Blob & any \\
    Access.dbInt & int & number \\
    Access.dbDouble & double & number \\  
    \end{tabular}%
  \label{tab:mappings-talbes}%
\end{table}%

\paragraph{Results}
\begin{table}[htbp]
  \centering
   \caption{DAO and DTO creation results. }
    \begin{tabular}{lrrrr}
    \textbf{Kind} &\textbf{Number} & \textbf{Java DTO} & \textbf{Java DAO} & \textbf{Typescript DTO} \\
    \midrule
    Tables & 20 & 0 error & 0 error &  0 error \\
    Queries & 27 &  27 error &  0 error & 0 error\\
    \end{tabular}%
  \label{tab:table-table-results}%
\end{table}%
\autoref{tab:table-table-results} shows the results of the experiment. 

On the \emph{Java} part of the experiment, we want to note that the queries generated by DTO were not compiling because we did not recognise the primary key of these tables, making it impossible to add the expected annotation javax.persistence.@Id to any attribute. 
We solved this problem by manually adding the annotations.

We tested each DAO manually, having successful results, which means that the code parses, compiles and executes as expected. 

On the \emph{Typescript} part of the experiment, we successfully produced all the required classes without any parsing or compiling errors. 

\paragraph{Discussion.}  
In our migration, we do not foresee mapping java persistence for using a full object model since migrating code is primarily used as records. 
This is why in the back end, we create one model entity per table or query (to be used as a data transfer object -- DTO) and a data access object (DAO) to list all the elements, save, update and delete. 
For the front end, we only need to define the DTO objects to be able to consume information from the back end. This is why we only produce classes. 

\subsubsection{Migrating a full Form: Form Login}
\label{sec:whole-migration}
\begin{figure}
	\centering
	\includegraphics[scale=0.30]{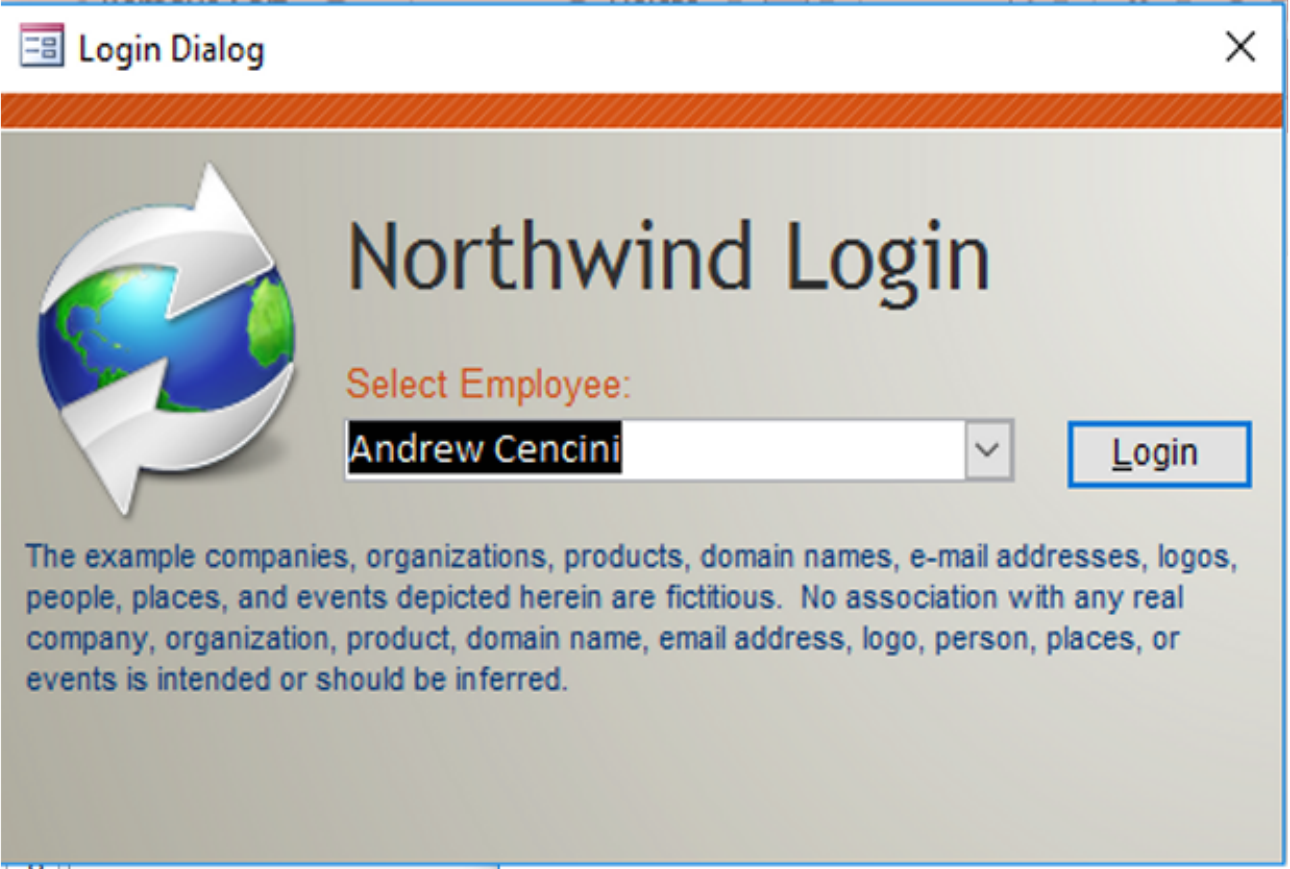}
	\caption{\label{fig:login-figure} Login northwind}
\end{figure}

We propose a validation that aims to give a holistic study case.
We fully migrated the login form MS Access Northwind Traders (depicted by \autoref{fig:login-figure}) to Java+Springboot and Typescript+Angular. 

This migration is simple, but it includes all the migrating aspects of our industrial case of migration: 
Splitting the application into the front end and the back end. 
Produce a back end that provides the data required by the front end. 
Produce a front end to consume data provided by the back end and show an equivalent UI. 

\paragraph{Rules} 
This experiment requires more \emph{Productive} rules than previous experiments and of different natures. 
\emph{Language translation:}  6 rules in the front end translating the source code of the event handling procedures. 
\emph{Data access migration:} 2 rules in the back end and 1 in the front end creating DAOs and DTOs like in the previous experiment.
\emph{Data requirement:} 3 rules in the back end and 1 in the front end creating SpringBoot and Angular services and controllers to interact based on the data required by the Form.
\emph{UI} 1 rule in the front end leveraging casino \cite{verh21b} that migrates the UI. Casino works from a GUI model that we generate by analysing the migrating form when the rule is applied.

We provide a single \emph{Adaptive} rule: SimpleRename in both the back and front end for migrating the types used.

\paragraph{Mappings} 
The mappings include all those defined in \autoref{tab:mappings-talbes} and

\paragraph{Results}

\begin{figure}
	\centering
	\includegraphics[scale=0.35]{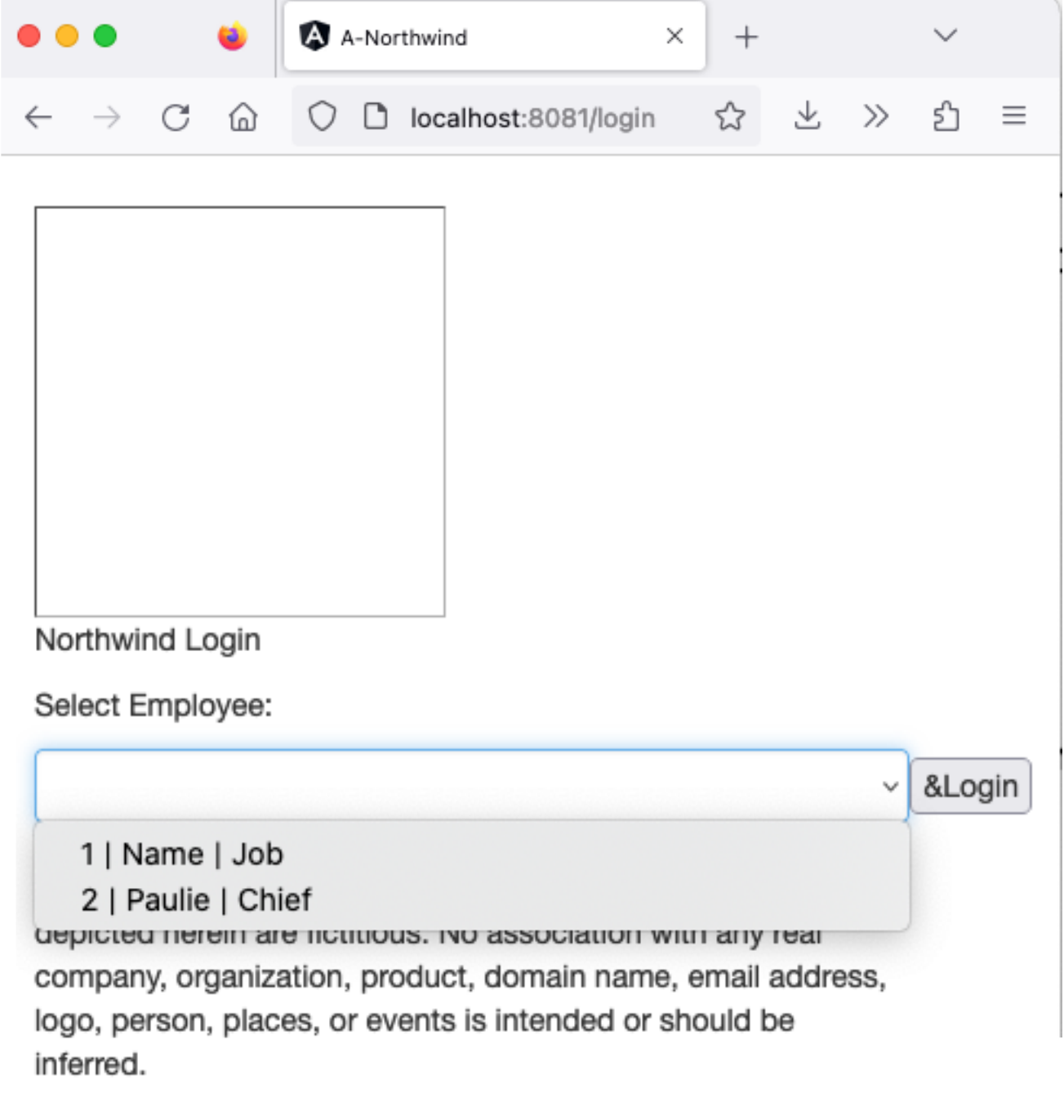}
	\caption{\label{fig:front-end-migrated} Generated Front End. The figure shows the visual aspect of the migrated version of the widget. It also shows that the dropdown list has been filled up with content from the backend. }
\end{figure}

\autoref{fig:front-end-migrated} depicts the migrated angular component shown in a browser. 

\begin{figure}
	\centering
	\includegraphics[scale=0.35]{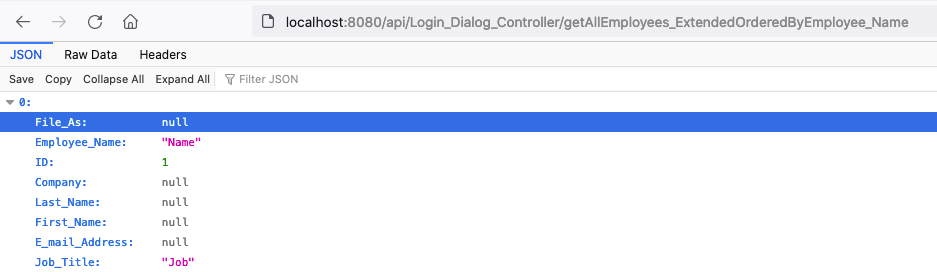}
	\caption{\label{fig:back-end-migrated} Generated Back-End API request. 
	The figure shows the manual invocation of the backend. 
	The endpoint called is the one required by the Form login to load the combo box. 
	The response is a JSON collection.}
\end{figure}

\autoref{fig:back-end-migrated} shows the result of manually executing the request for information to the backend. 

\paragraph{Making it work} 
The outcome of this example is not immediately compiling and working. 
We could achieve this by adding more rules. 
However, the missing rules would be extremely specific and hard to reuse. 
Therefore, we decided to go through the last mile by hand. 

The modifications done by hand are the following: 
\emph{Back end} Adding imports (some imports were not calculated), fixing the column names (in access names with special characters are allowed. Our test runs in a Postgres database, which does not allow this). 
\emph{Front end} Adding imports, configuring the angular modules and routing to add the component to the website. Changing a call to a method that does not exist by one that exists (to do it automatically would require a new adaptive rule); adding two functions required to fill up the HTML select control with the data coming from the back end: a function that extracts the ID from the DTO, and a function that gives a string to display from the same DTO. 
To make it work took approximately 30 minutes for one of the authors.

\paragraph{Discussion}

We can see that the component follows a similar layout to the original and has the same amount of components. 
It does not follow the same visual configuration since, for this experiment, we did not make any effort to migrate the visual design to CSS. 
However, we can access such information from the source model, and the Casino visual model is designed to handle it.

The backend includes a controller handling the login end-point. 
The implementation we provide now is just an exception announcing that a method must be implemented. 
In the source application, the click-handling procedure delegates the information to a library function. 
Our backend implementation handling is a method that responds to the same signature as the source library function but has no other behaviour than a runtime exception prompting the developer to implement its behaviour. 

\section{Discussion}
\label{sec:discussion}
\subsection{Programming paradigm migration}
\label{sec:paradigm-migration}

In the project, we must consider migrating from a procedural paradigm (VBA) to an Object-Oriented paradigm (Java, TypeScript or Pharo).
This is a complex issue that requires defining correct abstractions (classes) from variables and functions scattered in a procedural application  (\eg \cite{Dudd04a,Fox97a,Gros12a,Zou01}).
There is no well-accepted solution for this, and we chose to leave this issue to the migration tool user.
This means that the user is responsible for abstracting classes from the legacy application and migrating the relevant variables and functions to their appropriate classes.
As we saw in the preceding section, the tools assist in migrating functions to methods.
One tricky point remains for the invocation of methods which differ from function invocations in that they need a receiver.
We introduced three adaptive rules to handle some simple cases:
\begin{description}
\item[\rulename{RenameAdaptToStaticReceiver}:] This rule converts a function invocation to a method invocation using the class of the (static) method as the receiver. Presented in \autoref{sec:adaptiverules-static};
\item[\rulename{RenameAdaptToSameClassReceiver}:] This rule converts a function invocation to a method invocation using ``this'' or ``super'' as receiver, according to where the method is defined;
\item[\rulename{RenameAdaptToArgumentReceiver}:] This rule converts a function invocation to a method invocation using one argument of the function invocation as the receiver of the method invocation.
The execution of the rule includes asking the user which argument to use as the receiver.
\end{description}

One could think of a more complex rule (for example, one that would create an object on the spot), but it falls outside the scope of the current project.

\subsection{Rules}

\subsubsection{Rule morphology}

	Traditional model transformation approaches aim to be fully automatic (``push button'' approach), and reproducible \cite{Siik08a}.
	For this, the engine applies all the rules to every entity in the source model that it can handle.
	As such, transformation rules are typically composed of at least two parts (\eg \cite{Czar06a}): 
	(i) Left-Hand Side (LHS), a selection part specifying which source model element must be matched by the rule; and 
	(ii) Right Hand Side (RHS), a transformation part specifying what target model elements must be generated and how they should be initialised.
	The LHS part can be tricky to define in these approaches when one needs to precisely pinpoint a subset of entities, or a unique entity, for specific rules.

	
Our rules share this characteristic:
They have an LHS part (what we call condition) that filters the kind of entities they can apply to.
But because our rules are started by a user directive on a source entity and have a context of application, the filtering part is usually straightforward, for example, only checking for a specific entity type.

\subsubsection{Rule application}
	\paragraph{Reentrancy}
	A rule can always delegate its application over an entity's children.
	Like this, the rule that can create a Java method from an Access function will rely on other rules to create the parameters of the method or its body.
	 
	\paragraph{Scoping}
	Migration rules work from a source model to a target model.
	In our approach, the \emph{source model} of a rule is often, but not always, the legacy application model.
	Section \ref{sec:map-directive-application} gives examples of the source model of a rule being one of the migrated application models.
	Symmetrically, the \emph{target model} will usually be one of the migrated application models (actually, it is always the case in this paper). Still, we envision cases where the legacy application model could be the target of a transformation rule, for example, in a partial migration where parts of the legacy and migrated applications run together (see, for example, \cite{Verh22a}).
In such a case, the legacy application may need to be modified to reference some declaration node in the migrated application.

Rules are defined and applied within a \emph{context}.
The context can be the entire application, a specific package, a specific class,\ldots
Concretely, the context can be any declaration node in the model of an application \footnote{In the current implementation, we store the contexts in a ``twin tree'' of the target AST (i.e. with the same branches). For simplicity, this paper considers that contexts are stored in the AST.}.
When the context of the rule is the top-level node of an application model, this allows one to handle the specific programming language of this application.
For example, one could have an ``RPC''\footnote{Remote Procedure Call: When a method in a program calls a method defined in a different program, usually on another computer} package with special migration rules to add Spring Boot annotations automatically.

\subsection{Modelling}
	\subsubsection{Declarations and grammatical entities}
	The boundary between \emph{declaration} and \emph{grammatical} is not clear cut.
	It depends on whether we want to be able to handle references on an entity.
	For example, in Java, primitive types are part of the language's grammar. Still, they are modelled as declaration nodes because entities like variables have reference to them in their declarations and also because we might need to migrate them to other types (\eg Java boolean to C int).
Conversely, the ``+'' operator is currently modelled as a grammatical node. Still, it could be modelled as a function (a declaration node), for example, when this operator represents the string concatenation in Java.
In this case, the ``+'' in the Java expression `` \texttt{"the answer is:"+42} '' would be modelled as a reference to a particular java function \textsl{stringConcatenation}, with two arguments, so that one would be able to add migration rules for it.

We avoid modelling each programming language independently, trying to have, as much as possible, all language ASGs represented using the same types.
In this, we diverge from what a meta-model like Famix does (one specific meta-model for each programming language \cite{anqu20a}) and are closer to what KDM of the OMG proposes (one unique meta-model \cite{KDM11a}).
Our rationale for having only one meta-model is that when migrating from one language to another, we want to keep as much of the source ASG as possible.
Having the same nodes for both languages helps us do that as it allows us to ``migrate'' by just copying an entire branch of the ASG (e.g. a complex numerical expression) from the source model to the target one. This allowed us to use the same rules for this experiment when migrating to Pharo and Java. 

Yet, it is not always possible to do it because, as noted in \cite{anqu20a} ``various programming languages have minor semantic differences regarding how they implement [some programming concepts]''.
The same goes for ASG nodes; for example, some languages have an ``elseif''\footnote{For example, ``elseif'' in Visual Basic or PHP, ``elif'' in Python} statement (or part of the statement), and others represent it as an ``else'' containing a new ``if''.
(Note that KDM or GASTM do not offer an ``elseif'' node.)

To accurately migrate a language, these minor semantic differences must be modelled. Therefore, we use the same ASG nodes for different languages as long as the nodes do have the same semantics.
When this is not the case (\eg Java classes and TypeScript classes), we create different ASG nodes to be able to differentiate them in the migration rules.

\subsubsection{Model integrity} 
	
	Consequently, our meta-model favours permissiveness over semantic integrity. For example, we will allow all our ``if'' statements to have an ``elseif'' part even if this is not the case for a language.

	However, to be able to tell at any moment what entities in a given model are or are not correct, we reify the typing restrictions of a model concerning a given language by providing a Typing Ontology. 
	
	This device is used to give extra information to the user about what entities should be transformed in the target model to make it suitable to produce a compilable piece of code.
	
\section{Future work and Conclusion}
\label{sec:future-conclusion}

\subsection{Future work}

Our most probable direction is the development of a third directive: delegate. This directive is expected to propose the migration and further replacement of the source model to become an invocation of the freshly migrated piece of code. This would encourage and ease the idea of migration as a deflating process that takes the complexity out of the source.

During this work, we recognise many critical challenges involving the user experience, the thing that is chiefly important in the approach's productivity, such as:
(i) What visualisations would allow the developer to acquire \textbf{enough} information to take architectural decisions?
(ii) What visualisations would allow the developer to understand the immediate impact of applying a directive?
(iii) What kind of validation algorithm would inform the developer of the current problems of the migration?
(iv) Can migration rules be inferred from the UI interaction?
(v) How do we graphically debug rules?

We also got many process-related questions for the future, such as:
(i) How can we measure the process of migration? When does a migration finish?
(ii) Can we generate tests automatically knowing that we have a source implementation known to be correct?

\subsection{Conclusion}

A complete system migration is an overwhelming, complex and intimidating mission to achieve.

Our particular case of migration is a case with multiple layers of software migration, as we described in \autoref{sec:challenges}.
Translate language implies using different grammatical rules for saying the same and using entirely different constructions to think of problems when the paradigm behind them is distant.

Different languages grow surrounded by different communities addressing problems in different and often incompatible ways, favouring the apparition of different libraries and frameworks, from the definition of their concerns to the definition of their API.
Different deploying environments and architectural designs highlight different features and requirements impacting the inbound design in radically different ways, such as the differences between a monolithic application and a microservices one.
All these gaps impact at many levels; not only must the code change ultimately its nature:  the grammar, the APIs, the concepts and the concerns but the developers as well.
We claim that for large mission-critical projects with such a gap, fully automatic migration is almost as undesirable and unreliable as software rewriting since an automatically migrated mission-critical project would have no immediate maintainers nor developers at most of the aspects of the target projects: source code, library, architecture, building, shipping, and deploying. Even from the point of view of the database, which we do not migrate, the code interaction would change, meaning that the knowledge of the database administrators may fail with the new versions of the code. 

For tackling down, we propose an iterative, interactive migration process (\autoref{sec:overview-process})that is based on three central ideas: developer control, immediate feedback, and independent life cycle of source and target systems. Our proposal has as its major drawback the requirement of good migration planning. Architectural and design decisions have to be taken as in forward engineering. To promote experimentation (which is highly required for taking decisions), we reinforce the importance of not punishing bad decisions by being able to undo all modifications from both model and source-code points of view.

Our approach proposes to scope the kind of migrating approach within the design of each rule. 
In our migrating approach, language migration is compulsory as the intention is to abandon Microsoft Access completely. 
Therefore all our rules are based on white-box approaches. 

However, this lack of black-box or grey-box approach rules does not mean that the approach is incompatible with it. 
A simple and naive example of producing a wrapper is the rule applied when applying the produce directive with a controller method from the back end into a service class in the front end.

%
%
%
In this article, we presented a process of migration fully supported by an interactive UI, which resolves the user migrating interactions with a transformation engine based on directives.
For that, we introduced our infrastructure elements: directives, mappings/stubs, rules and multiple models (\autoref{sec:directive-rules}).
We explained step-by-step how these elements interact to achieve the user's desired activity (\autoref{sec:approach-in-action}) 

We validate the approach (\autoref{sec:valid}) by implementing the UI and transformation engine(\autoref{sec:implementation}) and applying it to three different cases of migration: 
We migrated library usage and paradigm invocations from Microsoft Access to Java and Pharo languages (\autoref{sec:lib-migration}).
We migrated the table and query objects from Microsoft Access to Java + Java persistence and Typescript (\autoref{sec:table-migration}).
We migrated a form and a query from Microsoft Access to Java+Springboot and Typescript+Angular, using the tool to create (i) the back end and the front end code and (ii) the required code to interact between the front end and back end (\autoref{sec:whole-migration}).

\bibliographystyle{alpha}
\bibliography{rmod,others,local}

\end{document}